\documentclass[11pt]{article}
\usepackage{amsmath}
\usepackage{amssymb}
\usepackage{amsthm}

\def\fpd#1#2{{\displaystyle\frac{\partial #1}{\partial #2}}}

\def\R{\mathbb{R}}

\def\onehalf{{\textstyle\frac12}}

\def\sode{{\sc sode}}

\newtheorem{prop}{Proposition}

\begin{document}

\title{Time-dependent kinetic energy metrics for Lagrangians of electromagnetic type}
\author{W.\ Sarlet$^{a,b}$, G.\ Prince$^b$, T.\ Mestdag$^a$ and O.\ Krupkov\'a$^{c,b}$\\[2mm]
{\small ${}^a$Department of Mathematics, Ghent University }\\
{\small Krijgslaan 281, B-9000 Ghent, Belgium}\\[1mm]
{\small ${}^b$Department of Mathematics and Statistics, La Trobe University}\\
{\small Bundoora, Victoria 3086, Australia}\\[1mm]
{\small ${}^c$Department of Mathematics, The University of Ostrava}\\
{\small 30.\ dubna 22, 70103 Ostrava, Czech Republic}
}

\date{}

\maketitle

\begin{quote}
{\small {\bf Abstract.} We extend the results obtained in \cite{SP10} about a class of Lagrangian systems which admit alternative kinetic energy metrics to second-order mechanical systems with explicit time-dependence. The main results are that a time-dependent alternative metric will have constant eigenvalues, and will give rise to a time-dependent coordinate transformation which partially decouples the system. }
\end{quote}

\section{Introduction}

In a recent contribution \cite{SP10} two of us examined the
following particular case of the inverse problem of Lagrangian
mechanics: suppose that an autonomous second-order system admits a
Lagrangian description such that the Hessian of the Lagrangian $L$
is the unit matrix; what are the conditions for the existence of an
alternative kinetic energy metric for the same system, or expressed
differently, an alternative multiplier for the inverse problem,
which depends on the position variables only. Two observations we
made are worth mentioning right away: (i) the starting assumption
implies that the forces under consideration necessarily are of
(generalized) electromagnetic type; (ii) if a second multiplier
exists, it necessarily will have to be constant. We obtained several
interesting results for such systems, one of them being that they
necessarily decouple in coordinates which diagonalize the
alternative metric.

It turns out to be quite interesting to enlarge the scope of
applicability of the preceding study by allowing explicit
time-dependence into the picture. This is the subject of the present
paper. We will see indeed that such an extension leads to
non-trivial complications, such as the fact that the determination
of admissible scalar and vector potentials is no longer a separate
issue. Nevertheless, the conclusion about partial decoupling of the
system in appropriate coordinates will be shown to remain intact.

The literature about the inverse problem of Lagrangian dynamics is
vast. For a small sample of different techniques and mathematical
tools which have been used, see \cite{S82, AT92, Olga, GM1, APST,
Buca} and the list of references in the review paper
\cite{olgageoff}. A number of these techniques require quite
advanced methods of differential geometry. Note that allowing
explicit time-dependence in the framework is not always a
straightforward matter; it involves, for example, passing from the
vector bundle structure of a tangent bundle to the affine context of
a first jet bundle. For specific aspects of time-dependence in this
context, see for example \cite{Buca2, MP, olgageoff, SVCM95}. The
very concrete problem we are addressing here is much more
computational in nature, so that it will be sufficient to rely on
analytical expressions of the conditions for the existence of a
Lagrangian, such as the ones developed in \cite{S82}.

The second-order systems (\sode s for short) we start from are
general non-autonomous differential equations in normal form, say
\begin{equation}
\ddot{x}^a = F^a(t,x,\dot{x}), \qquad a=1,\ldots, n.
\label{diffeqns}
\end{equation}
Geometrically they are the analytical expression of a second-order
equation field $\Gamma$ living on the first jet bundle $J^1\pi$ of a
bundle $\pi:E\rightarrow \R$, so
\[
\Gamma = \fpd{}{t} + \dot{x}^a \fpd{}{x^a} +
F^a(t,x,\dot{x})\fpd{}{\dot{x}^a}.
\]
The local formulation of the general inverse problem is the question
for existence of a (non-singular) multiplier matrix $g_{ab} (t,x,
\dot x)$, such that
\[
g_{ab} (\ddot x^b - F^b) \equiv \frac{d}{dt} \left(\frac{\partial
L}{
\partial \dot x^a}\right) - \frac{\partial L}{\partial x^a},
\]
for some Lagrangian function $L(t,x,\dot x)$. The necessary and
sufficient conditions for the existence of $L$ are generally
referred to as the {\it Helmholtz conditions}, and when expressed in
terms of conditions on the multiplier matrix, they read (see
\cite{S82}):
\begin{equation}
g_{ab} = g_{ba}, \quad \Gamma(g_{ab}) =
g_{ac}\Gamma_b^c+g_{bc}\Gamma_a^c, \quad g_{ac}\Phi_b^c  =
g_{bc}\Phi_a^c, \quad \frac{\partial g_{ab}}{\partial\dot x^c}  =
\frac{\partial g_{ac}}{\partial\dot x^b}. \label{Helmholtz1}
\end{equation}
Here
\[
\Gamma_b^a := -{\frac{1}{2}}\frac{\partial F^a}{\partial \dot x^b}
\]
geometrically represent the components of the canonical non-linear
connection associated to $\Gamma$, so they are non-tensorial
objects. On the other hand,
\[
\Phi_b^a := -\frac{\partial F^a}{\partial x^b} -
\Gamma_b^c\Gamma_c^a - \Gamma(\Gamma_b^a)
\]
are the components of a type $(1,1)$ tensor field along the
projection $\pi^1_0: J^1\pi \rightarrow E$ called the Jacobi
endomorphism. For a systematic development of the geometrical
calculus along $\pi^1_0$, the interested reader can consult
\cite{SVCM95}. Further worth mentioning for later reference is that
an integrability study of the conditions for the $g_{ab}$ leads to
more algebraic restrictions, the first one of which, called the
\emph{curvature condition\/}, reads
\begin{equation}
g_{ar}R^r_{bc} + g_{br}R^r_{ca} + g_{cr}R^r_{ab} = 0. \label{R}
\end{equation}
Here,
\[
R^a_{bc}:= H_c(\Gamma^a_b) - H_b(\Gamma^a_c), \quad \mbox{where}
\quad H_c = \fpd{}{x^c} - \Gamma_c^r\fpd{}{\dot{x}^r}.
\]
are components of the curvature tensor of the non-linear connection
referred to before.

The concrete question we shall address in this paper is the
characterization of (generally time-dependent) second-order systems
which have a Lagrangian representation with a Euclidean kinetic
energy term, and which admit at least one alternative Lagrangian
representation for which the multiplier matrix does not depend on
the velocities but may explicitly depend on time. The main
conditions to be satisfied by such systems are developed in
section~2; one of the conclusions is that the alternative multiplier
$g$ can in fact be a function of time only. In section~3, it is
shown that this $g$ will have constant eigenvalues and that a linear
coordinate transformation which diagonalizes $g$ will decouple the
system into lower dimensional Lagrangian systems. Explicit examples
which illustrate the various aspects of the theory are presented in
section~4. In section~5, we are led to an additional
characterization of the systems under consideration, in the
particular case that the connection coefficients are functions of
time only.

\section{Lagrangians of electromagnetic type and alternative multipliers}

Suppose $\Gamma$ is a \sode\ field on $J^1\pi$ with the property
that, in some coordinate system, the unit matrix:
$g_{ab}=\delta_{ab}$ is a multiplier matrix for the inverse problem.
In fact, for simplicity, we shall take $\pi$ to be the trivial
bundle $\pi:\R\times\R^n \rightarrow \R$. Since the unit matrix is
the Hessian of the function $\onehalf \sum (\dot{x}^a)^2$, our
assumption means that there exists a Lagrangian $L$ for the given
\sode, which differs from that quadratic function at most by a
linear function in the velocities, say  $L=\onehalf
\sum(\dot{x}^a)^2 + A_c(t,x)\dot{x}^c - V(x,t)$. This implies that
the `forces' will be of the form ($A^a=\delta^{ab}A_b$)
\begin{equation}
F^a = \left(\fpd{A^b}{x^a}-\fpd{A^a}{x^b} \right)\dot{x}^b -
\fpd{V}{x^a} - \fpd{A^a}{t}. \label{forces}
\end{equation}
It further follows that the connection coefficients are
skew-symmetric:
\begin{equation}
\Gamma^a_b := - \onehalf \fpd{F^a}{\dot{x}^b} =
\frac{1}{2}\left(\fpd{A^a}{x^b}-\fpd{A^b}{x^a}\right),
\label{conncoeff}
\end{equation}
while the components of the Jacobi endomorphism become
\begin{equation}
\Phi^a_c = \frac{1}{2}\left(\fpd{^2A^a}{x^c\partial x^b} +
\fpd{^2A^c}{x^a\partial x^b} - 2\fpd{^2A^b}{x^a\partial x^c}\right)
\dot{x}^b - \Gamma^a_b\Gamma^b_c + \fpd{^2V}{x^a\partial x^c} +
\onehalf\left(\fpd{^2A^a}{t\partial x^c} + \fpd{^2A^c}{t\partial
x^a}\right) \label{Phi}
\end{equation}
and are seen to be symmetric.

We want to investigate under what conditions there exists a second
multiplier for such systems, which does not depend on the
velocities. If $g$ does not depend on the velocities it follows from
the requirement
\[
\Gamma(g_{ab})= \fpd{g_{ab}}{t}+\fpd{g_{ab}}{x^c}\dot{x}^c =
g_{ac}\Gamma^c_b + g_{bc}\Gamma^c_a,
\]
that $g$ must actually be a function of $t$ only for which it must
further hold that
\begin{equation}
\dot{g}_{ab}(t) = g_{ac}(t)\Gamma^c_b + g_{bc}(t)\Gamma^c_a.
\label{dotg}
\end{equation}
Notice that this implies
\begin{equation}
g_{ac}(t)\fpd{\Gamma^c_b}{x^r} + g_{bc}(t)\fpd{\Gamma^c_a}{x^r}=0,
\label{skewderiv}
\end{equation}
which is a somewhat weaker skew-symmetry property than in the
autonomous case. The remaining Helmholtz condition $g_{ac}\Phi^c_b =
g_{bc}\Phi^c_a$ is a linear expression in the $\dot{x}^a$ as well
and thus also splits into two conditions. From the terms which are
linear in the velocities it can be shown that the weaker condition
(\ref{skewderiv}) is sufficient to imply that (as in the autonomous
case) we still must have
\begin{equation}
\fpd{\gamma_{ar}}{x^b} + \fpd{\gamma_{rb}}{x^a} +
\fpd{\gamma_{ba}}{x^r} =0, \label{curv}
\end{equation}
with $\gamma_{ab}:=g_{ac}\Gamma^c_b$. This may surprise at first
sight, but it is fully consistent with the curvature condition
(\ref{R}) for this case. The remaining terms in the
$\Phi$-condition, which in the autonomous case produced conditions
on admissible scalar potentials $V$ only, now give rise to the
following more complicated requirements, involving both the scalar
and vector potential:
\begin{eqnarray}
\lefteqn{ g_{ac}\left[\fpd{^2V}{x^c\partial x^b} +
\onehalf\left(\fpd{^2A^c}{t\partial x^b} + \fpd{^2A^b}{t\partial
x^c}\right) - \Gamma^c_r\Gamma^r_b\right] = } \nonumber \\
&& \hspace{1cm} g_{bc}\left[\fpd{^2V}{x^c\partial x^a} +
\onehalf\left(\fpd{^2A^c}{t\partial x^a} + \fpd{^2A^a}{t\partial
x^c}\right) - \Gamma^c_r\Gamma^r_a\right]. \label{Veqn}
\end{eqnarray}

Summarizing, we come to the following conclusion.

\begin{prop} Assume that the \sode\ $\Gamma$ on $\R^{2n+1}$ admits a coordinate
representation for which $\delta_{ab}$ is a multiplier matrix for
the inverse problem of Lagrangian mechanics. Then the right-hand
sides of the equations are of the form (\ref{forces}) for some
functions $A^a$ and $V$. Any admitted alternative multiplier
$g_{ab}$, not a multiple of the identity and independent of the
velocities, must be a function of \,$t$ only, must satisfy the
differential conditions (\ref{dotg}), the curvature type conditions
(\ref{curv}) with respect to the functions $A^a$, and the conditions
(\ref{Veqn}) with respect to the functions $V$ and $A^a$. \qed
\end{prop}

In the next section, we shall see that these conditions imply
existence of a linear change of coordinates which partially
decouples the given equations.

An interesting special case which complements the results of
\cite{SP10} is when the system with given forces (\ref{forces}) does
depend explicitly on time, but we restrict the search for an
alternative multiplier to a constant $g$. It then follows from
(\ref{dotg}) that the matrix $g\Gamma$ must be skew-symmetric. As a
result the terms with two factors $\Gamma$ in the condition
(\ref{Veqn}) cancel out, but compared to the autonomous case in
\cite{SP10} there remains a coupling between the scalar and vector
potential. From the fact that also $g\partial \Gamma/\partial t$
will be skew-symmetric then, the remaining condition (\ref{Veqn})
can be further manipulated to reduce to
\[
g_{ac}\fpd{}{x^b}\left(\fpd{V}{x^c}+\fpd{A^c}{t}\right) =
g_{bc}\fpd{}{x^c}\left(\fpd{V}{x^a}+\fpd{A^a}{t}\right).
\]

\section{Diagonalization of $g$ and decoupling}

We observe that the differential conditions (\ref{dotg}) for $g$
have the form of a Lax equation because the matrix $\Gamma^a_b$ is
skew-symmetric. It follows that the eigenvalues of $g$ are constant.
The point now is that the diagonalization of $g$ can be achieved by
a linear coordinate change and that the connection coefficients,
although not behaving tensorially under such operation, transform
exactly in the way which is needed to contribute to decoupling of
the system.

Let $P(t)$ be an orthogonal matrix which diagonalizes the symmetric
matrix $g(t)$ and put $\bar{g}= P^TgP$, so that $\bar{g}$ is
diagonal. Using an obvious matrix notation, the equation
(\ref{dotg}) can be written as
\[
\dot{g} = g\Gamma - \Gamma g.
\]
[It will always be clear from the context whether the notation
$\Gamma$ refers to the \sode\ vector field or to the matrix of
connection coefficients.] Multiplying on the left by $P^T$ and on
the right by $P$, the equation transforms to
\begin{equation}
\dot{\bar{g}} = \bar{g}(\bar{\Gamma}+ P^T\dot{P}) - (\bar{\Gamma}+
P^T\dot{P})\bar{g}, \label{diagg}
\end{equation}
where $\bar{\Gamma}=P^T\Gamma P$. Notice that $\bar{\Gamma}$ is
still skew-symmetric and the same is true for
$\widetilde{\Gamma}:=\bar{\Gamma} + P^T\dot{P}$. It trivially
follows from the fact that $\bar{g}$ is constant that the commutator
in the right-hand side is zero.

Now consider the linear, time-dependent coordinate transformation
$x=P(t)y$ with inverse $y=P(t)^Tx$. Both $g_{ab}$ and $\Phi^a_b$ are
actually components of tensor fields along $\pi^1_0:J^1\pi
\rightarrow E$ of the form (see \cite{SVCM95})
\[
g = g_{ab}\theta^a\otimes\theta^b, \qquad \Phi=\Phi^a_b\theta^b
\otimes \fpd{}{x^a} \qquad \mbox{with }\quad \theta^a=dx^a-\dot{x}^a
dt.
\]
Hence they behave well under time-dependent coordinate
transformations, meaning that their components in the new variables
under consideration will be $\bar{g}$ and $\bar{\Phi}=P^T\Phi P$.
Less obvious is what happens to the non-tensorial matrix of
connection coefficients. It is a straightforward computation,
however, to verify that the transformed second-order differential
equations for the $y$-variables contain the following linear part in
the velocities:
\[
\ddot{y} = -2(P^T\Gamma P - \dot{P}^TP)\dot{y} + \ldots.
\]
Thus, the connection coefficients $\widetilde{\Gamma}^a_b$ of the
transformed equations are precisely the components of the matrix
$\widetilde{\Gamma}$ introduced above. Since $\widetilde{\Gamma}$
commutes with the diagonal matrix $\bar{g}$, it has a block diagonal
structure, the dimension of each block being determined by the
multiplicity of an eigenvalue of $g$. It follows that, if $A_\alpha$
and $B_\beta$ refer to coordinates of different blocks, the
transformed forces $\widetilde{F}$ will have the property $\partial
\widetilde{F}^{A_\alpha}/\partial \dot{y}^{B_\beta}=0$. But we also
have $\bar{g}\bar{\Phi} = \bar{\Phi}\bar{g}$, meaning that
$\bar{\Phi}$ has the same block diagonal structure, and from the
coordinate expression of $\bar{\Phi}^a_b$ (see (\ref{Phi}), but to
be read in the new variables), it then further follows that also
$\partial \widetilde{F}^{A_\alpha}/\partial y^{B_\beta}=0$. Hence,
the transformed equations decouple into independent subsystems, one
for each eigenspace of $g$.

\begin{prop}
Assume that a given \sode\ $\Gamma$ satisfies the conditions of
proposition~1. Then the linear time-dependent coordinate
transformation which diagonalizes the second multiplier $g_{ab}(t)$
will decouple the system into a number of separate \sode s, one for
each distinct eigenvalue of $g$, and these eigenvalues are
constants. \qed
\end{prop}

As we explained in \cite{SP10} such a decoupling property was
already established, for autonomous systems, in work of Ferrario et
al \cite{Ferrario}, but to the best of our knowledge, it was not
observed before that allowing time-dependence in the picture does
not destroy it. Observe also that the transformed system (and hence
every independent subsystem) still has the property of admitting the
unit matrix as multiplier, because $P^T I_nP= I_n$. As a result, a
reverse engineering technique to construct all systems under
consideration goes as follows: start from a number of say $k$
independent \sode s with forces of type (\ref{forces}); multiply
each of them with a different constant factor $\lambda_k$, thus
creating a diagonal multiplier $\bar{g}$ for the union of the $k$
systems; then apply an arbitrary (possibly time-dependent) linear
orthogonal transformation to the overall system; the resulting
coupled system will be of the type characterized in Proposition~1.
Note that each choice of a single equation at the start of this
process, corresponding to an eigenvalue with multiplicity one of the
$g$ under construction, will only contain a scalar potential. Since
the group of orthogonal transformations is finitely generated, one
can imagine that such a reverse engineering technique may be helpful
in an attempt to classify all systems in the category under
consideration. But there would of course, in such a classification,
be room for an unlimited number of ways of putting together
decoupled systems in the first place.

The more interesting approach for identifying suitable examples,
therefore, is to start from an arbitrary \sode\ with forces
(\ref{forces}) and to try to solve the conditions of Proposition~1
for the identification of admissible scalar and vector potentials
and the construction of an associated alternative (possibly
time-dependent) kinetic energy function. This is the line of
approach we will adopt as much as possible in the next section.

\section{Examples}

Consider a system $\ddot{x}^a=F^a(x,\dot{x},t)$, where the $F^a$ are
of the form (\ref{forces}). In the autonomous case we were able to
obtain relevant information about the form of admissible vector
potentials and the alternative multiplier $g$ by imposing first that
the matrix $\gamma:=g\Gamma$ had to be skew and satisfy the
curvature type condition (\ref{curv}). The remaining Helmholtz
condition $g\Phi=(g\Phi)^T$ then resulted essentially in equations
for the admissible scalar potential $V$. The situation is more
complicated in the time-dependent case now, for the following three
reasons: firstly, $\gamma$ need not be skew, instead, in view of
(\ref{skewderiv}), we have the weaker condition that $\gamma -
\gamma^T$ must be a function of time only; secondly, the information
about $g$ which can be gathered in the process of analysing $\gamma$
will be far less conclusive, because $g$ still has to satisfy the
differential equation (\ref{dotg}) in the end; thirdly, the
remaining algebraic condition (\ref{Veqn}) will in general not
reduce to a separate issue concerning admissible functions $V$. As
we shall show below, however, we can still get a long way in
constructing examples for $n=2$ and $n=3$ by staying close to the
procedure we followed in the autonomous case.

Taking $n=2$, and denoting the elements of the multiplier $g$ for
shorthand by $a_1=g_{11},\ a_2=g_{22}, \ b=g_{12}=g_{21}$, the fact
that $\gamma - \gamma^T$ can only depend on $t$ implies that
$b\,\Gamma^1_2$  and $(a_1-a_2)\Gamma^1_2$ must be functions of time
only. It follows that $\Gamma^1_2$ cannot depend on the
$x$-coordinates either (which we write with lower indices here);
putting $\Gamma^1_2=\sigma(t)$, the vector potential must satisfy
\[
\fpd{A_1}{x_2} - \fpd{A_2}{x_1} = 2\sigma(t),
\]
and is determined, up to a gradient, by
\begin{equation}
A_1(t,x) = \sigma(t)\,x_2, \qquad A_2(t,x) = - \sigma(t)\,x_1.
\label{An2}
\end{equation}
The freedom of adding a gradient is irrelevant here since this adds
to the Lagrangian $L$ we started from merely a total-time derivative
term, plus a term which can be absorbed into the as yet undetermined
scalar potential $V$. Expressed differently, we can assume without
loss of generality that in dimension $n=2$ the system we start from
can be written in the form
\begin{eqnarray*}
\ddot{x}_1 &=& - 2\sigma(t)\dot{x}_2 - \dot{\sigma}x_2 -
\fpd{V}{x_1}, \\
\ddot{x}_2 &=& 2\sigma(t)\dot{x}_1 + \dot{\sigma}x_1 - \fpd{V}{x_2}.
\end{eqnarray*}
Quite remarkably, if we now move to the remaining algebraic
condition (\ref{Veqn}), all terms not involving $V$ cancel out and
we end up with the following condition for admissible functions $V$:
\begin{equation}
b\left(\fpd{^2V}{x_2^2} - \fpd{^2V}{x_1^2}\right) =
(a_2-a_1)\fpd{^2V}{x_1\partial x_2}.
\end{equation}
By way of example we proceed to determine admissible
(time-dependent) quadratic  potentials. Putting
\[
V=\onehalf k(t) x_1^2 + l(t) x_1x_2 + \onehalf m(t) x_2^2,
\]
the condition becomes
\[
b(m-k) = (a_2-a_1)\,l
\]
and we will only discuss the generic case here that $l\neq 0$ and
$m\neq k$.

Putting $l(t)=(m(t)-k(t))\rho(t)$ the multiplier must have the form
\[
g= \left( \begin{array}{cc} a_1 & (a_2-a_1)\,\rho \\ (a_2-a_1)\,\rho
& a_2
\end{array}\right)
\]
and the corresponding admissible potentials read
\begin{equation}
V=\onehalf k x_1^2 + (m-k)\rho\, x_1x_2 + \onehalf m x_2^2.
\label{V2}
\end{equation}
But now $g$ still must satisfy the differential equations
(\ref{dotg}). Putting $a_2= C - a_1$, where $C$ represents the
constant trace of $g$, these equations reduce to:
\begin{eqnarray*}
\dot{a}_1 &=& 2(2a_1-C)\rho\sigma \\
\dot{\rho} &=& - (1+4\rho^2)\sigma.
\end{eqnarray*}
The case $2a_1-C=0$ has to be excluded since $g$ must be kept
different from a multiple of the identity. It is convenient to
introduce a new time-scale $\tau:=\int \sigma(t)dt$; the equations
then are easy to integrate, giving
\[
\rho(\tau) = \onehalf \tan(-2\tau + A), \qquad 2a_1(\tau)-C = B \cos
(-2\tau + A),
\]
where $A$ and $B$ are constants.

In conclusion, with $k(t)$, $l(t)$, $\sigma(t)$ arbitrary functions
of time, $\alpha(t)= -2\int^t\sigma(s)ds+A$ and $\rho(t)=\onehalf
\tan\alpha$, the admissible scalar and vector potentials are given
by (\ref{V2}) and (\ref{An2}) respectively, and the alternative
multiplier is
\[
g= \left( \begin{array}{cc} \onehalf(C+B\cos\alpha) & -\onehalf
B\sin\alpha \\-\onehalf B\sin\alpha & \onehalf(C-B\cos\alpha)
\end{array}\right).
\]
The eigenvalues of $g$ are $\lambda_1=\onehalf(C-B), \lambda_2=
\onehalf(C+B)$ and corresponding eigenvectors read $(\sin\alpha,
\cos\alpha+1),\ (\sin\alpha, \cos\alpha-1)$. They have to be scaled
to norm 1 to produce the columns of an orthogonal matrix which
diagonalizes $g$. For testing the statement about decoupling in
Proposition~2, however, one can relax the requirement that $P$ be
orthogonal. Indeed, if we look at the related coordinate
transformation in the form $y=P^T x$ whereby the eigenvectors now
appear as rows, it is clear that an overall factor in the defining
relation of the $y$-variables cannot be an obstruction to
decoupling. One can verify that the transformation
\begin{eqnarray*}
y_1 &=& \sin{\alpha}\, x_1 + (\cos\alpha +1) x_2 \\
y_2 &=& \sin\alpha\, x_1 + (\cos\alpha -1) x_2,
\end{eqnarray*}
effectively takes care of that.

Let's move now to systems with three degrees of freedom, for which
the problem is more intricate because the curvature condition
(\ref{curv}) comes into play. To better illustrate the effect of
allowing time-dependence we try to stay close here to the analysis
and notations of the $n=3$ example in \cite{SP10}. As indicated
before, the matrix $\gamma=g\Gamma$ need not be skew-symmetric now,
but its symmetric part, according to (\ref{skewderiv}) can depend on
time only. In particular, from the diagonal elements of this
symmetric part, the following expressions (which had to be zero in
the autonomous case) now must be certain functions of time (which we
will not specify at the moment):
\[
g_{12}\Gamma^1_2 + g_{13}\Gamma^1_3 , \quad g_{12}\Gamma^1_2 -
g_{23}\Gamma^2_3 , \quad g_{13}\Gamma^1_3 + g_{23}\Gamma^2_3,
\]
and we can view this as a linear system for the $\Gamma^a_b(t,x)$
(with time-dependent coefficients) which has zero determinant.
Except for special cases where the rank of the coefficient matrix is
lower than two (and which we will not discuss), solving this
algebraically for the $\Gamma^a_b$ will give rise to relations of
the type
\begin{equation}
\Gamma^1_3= a(t)\,\Gamma^1_2 + \tilde{a}(t), \qquad \Gamma^2_3=
b(t)\,\Gamma^1_2+ \tilde{b}(t), \label{Gamma13-23}
\end{equation}
for certain functions $a, b, \tilde{a}, \tilde{b}$. The sort of
privileged role which $\Gamma^1_2(t,x)$ plays in these expressions
is just a matter of renumbering the variables, if necessary.
Substituting this structural information back into the linear
equations from which it was obtained, we see that the product of
$\Gamma^1_2(t,x)$ with any of the factors
\[
g_{12}+a g_{13}, \quad g_{12} - b g_{23}, \quad ag_{13} + bg_{23},
\]
must be some function of time only. There are then two cases to be
considered: either $\Gamma^1_2$ also depends on time only, and then
the same is true for $\Gamma^1_3$ and $\Gamma^2_3$; or we insist
that $\Gamma^1_2$ should depend on at least one of the
$x$-coordinates, and then the above mentioned factors necessarily
must be zero. We investigate the second option first.

With $\Gamma^1_3$ and $\Gamma^2_3$ as specified above, the off
diagonal elements of the symmetric part of $\gamma$ give us three
more algebraic conditions, which are affine linear in $\Gamma^1_2$.
Since we assume $x$-dependence in $\Gamma^1_2$, such relations can
hold true only if the coefficients of $\Gamma^1_2$ vanish which
yields
\[
g_{11}-g_{22} = b g_{13} + ag_{23}, \quad a(g_{11}-g_{33}) = g_{23}
- b g_{12}, \quad b(g_{22}-g_{33}) = -ag_{12} - g_{13}.
\]
We now have sufficient elements to get an idea of the possible
structure of the multiplier $g$. We must have
$g_{12}=-ag_{13}=bg_{23}$ and to make $ag_{13} + bg_{23}$ zero as
well, we can put $g_{13}= - \rho b, g_{23}= \rho a$, which implies
that $g_{12}=\rho a b$. Putting further $g_{11}=\rho c$ (which
defines the function $c(t)$), the last three conditions fix $g_{22}$
and $g_{33}$. The result is that the multiplier $g$ will have the
structure:
\begin{equation}
g=\rho \left( \begin{array}{ccc} c & ab & -b \\ ab & c+b^2-a^2 & a \\
-b & a & c+b^2 -1 \end{array} \right), \label{gforn=3}
\end{equation}
where $a,b,c$ and $\rho$ are all functions of time which will have
to be determined from the differential equations which $g$ must
satisfy. It is instructive at this stage to compare with the
situation in the autonomous case (see \cite{SP10}). It is not true
here that if $g$ is a multiplier matrix, then $\rho(t)g$ is one too,
so that the overall factor cannot be disregarded right away.

In dimension three the curvature type condition contains one
requirement and with the information gathered so far, it can be seen
to reduce to
\[
\fpd{\Gamma^1_2}{x_3} - a\, \fpd{\Gamma^1_2}{x_2} +
b\,\fpd{\Gamma^1_2}{x_1} = 0,
\]
which is formally the same as in the autonomous case. The fact that
the coefficients depend on time now is not an obstruction to take
over suitably adapted results we know from the autonomous situation.
The general solution of the above linear partial differential
equation is an arbitrary function $\Gamma^1_2=f(t,u,v)$ where
\begin{equation}
 u(t,x) = x_1 - b(t)\,x_3, \qquad v(t,x)= x_2 + a(t)\,x_3. \label{uv}
\end{equation}
This means that admissible vector potentials must satisfy
\begin{eqnarray*}
\fpd{A_1}{x_2} - \fpd{A_2}{x_1} &=& 2\,f(t,u,v), \\
\fpd{A_1}{x_3} - \fpd{A_3}{x_1} &=& 2a(t)\,f(t,u,v) + 2\tilde{a}(t), \\
\fpd{A_2}{x_3} - \fpd{A_3}{x_2} &=& 2b(t)\,f(t,u,v)+ 2\tilde{b}(t).
\end{eqnarray*}
The right-hand sides are the components of a closed 2-form indeed
(with $t$ treated as parameter), and a particular solution for the
$A_i$ then is given by
\begin{eqnarray}
A_1 &=& 2 v \int_0^1 s f(t,su,sv)ds + \tilde{a}(t)x_3, \nonumber \\
A_2 &=& -2 u \int_0^1 s f(t,su,sv)ds + \tilde{b}(t)x_3, \label{A} \\
A_3 &=& -2 (au+bv)\int_0^1 s f(t,su,sv)ds - \tilde{a}(t)x_1 -
\tilde{b}(t)x_2. \nonumber
\end{eqnarray}
The general solution differs from this particular one by a gradient,
but as argued in the preceding example, this gradient term can be
incorporated into the as yet undetermined scalar potential, up to a
total time derivative. We have obtained valuable information about
the structure of $g$ and the vector potential now, but it is clear
that the freedom in selecting functions $a,b,c$ and $\rho$ will be
severely restricted if we now go back to the differential conditions
(\ref{dotg}). They read:
\begin{eqnarray*}
&& \hspace*{-2cm}\frac{d}{dt}(\rho c) = 2\rho\tilde{a}b, \\
&& \hspace*{-2cm}\frac{d}{dt}(\rho(c+b^2-a^2)) = -2\rho\tilde{b}a, \\
&& \hspace*{-2cm}\frac{d}{dt}(\rho(c+b^2)) = 2\rho(\tilde{b}a-\tilde{a}b), \\
&& \hspace*{-2cm}\frac{d}{dt}(\rho ab) = \rho(b\tilde{b}-a\tilde{a}), \\
&& \hspace*{-2cm}\frac{d}{dt}(\rho a) = \rho\big(\tilde{b}(1-a^2) + \tilde{a}ab\big), \\
&& \hspace*{-2cm}\frac{d}{dt}(-\rho b) = \rho\big(\tilde{a}(1-b^2) +
\tilde{b}ab\big).
\end{eqnarray*}
It follows from the second and third equation that we must have
\[
\frac{d}{dt}(\rho a^2) = 2\rho(-\tilde{a}b + 2\tilde{b}a).
\]
Combining this with information coming from the fifth and sixth
equation, one can derive the condition
\[
a\tilde{b}-b\tilde{a}=0,
\]
and subsequently that
\[
\dot{a}=\tilde{b} \qquad\mbox{and}\qquad \dot{b}=-\tilde{a},
\]
which in turn implies that $a^2+b^2$ is constant, and from the third
equation also that $\rho(c+b^2)$ is constant. Finally, going back to
the fifth and sixth equation, which reduce to
\[
\frac{d}{dt}(\rho a) = \rho\tilde{b}, \qquad \frac{d}{dt}(\rho b) =
-\rho\tilde{a},
\]
and using the equations for $a$ and $b$ just obtained, it follows
that actually also $\rho$ must be constant and can therefore, from
now on, without loss of generality be put equal to~1. With these
restrictions, all six equations about $\dot{g}$ are now satisfied.
We conclude that the freedom in constructing alternative multipliers
consists of two arbitrary constants and one arbitrary function of
time. Take $a(t)$ to be arbitrary for example, and choose two
constants $c_1$ and $c_2$, then define $b$ by $a^2+b^2=c_1$ and $c$
by $c+b^2=c_2$. This fixes $g$, and the corresponding admissible
vector potentials are now also determined by the relations
(\ref{A}), where $f(t,u,v)$ is an arbitrary function of the
indicated variables and $\tilde{a}(t)=-\dot{b}$,
$\tilde{b}(t)=\dot{a}$.

It remains to find the form of the admissible scalar potentials for
which purpose we will avoid the conditions (\ref{Veqn}) which are
quite complicated to handle. Observe first that $g$ (as in
(\ref{gforn=3}) with $\rho=1$) has a double eigenvalue
$\lambda=c+b^2=c_2$ and a single eigenvalue $\mu=c-a^2-1=c_2-c_1-1$.
From a different perspective, the solution we obtained for
admissible vector potentials more or less forces a kind of
privileged coordinate transformation upon us. Indeed, extending
(\ref{uv}), consider the coordinate transformation (similar to the
autonomous case in \cite{SP10})
\begin{eqnarray}
u &=& x_1 - b(t)x_3, \nonumber \\
v &=& x_2 + a(t)x_3, \label{uvz} \\
z &=& x_3 + b(t)x_1 - a(t)x_2. \nonumber
\end{eqnarray}
The inverse transformation reads
\begin{eqnarray}
x_1 &=& k^{-1}\big((a^2+1)\,u + ab\,v + b\,z\big), \nonumber \\
x_2 &=& k^{-1}\big(ab\,u + (b^2+1)\,v - a\,z\big), \label{inverse} \\
x_3 &=& k^{-1}(-b\,u + a\,v + z), \nonumber
\end{eqnarray}
where $k=a^2+b^2+1$. If $P(t)$ denotes the coefficient matrix in
these inverse relations, it is easy to verify that its first two
columns are eigenvectors of $g$ corresponding to the eigenvalue
$\lambda$, and the third column is an eigenvector for the eigenvalue
$\mu$. Although this $P$ is not an orthogonal matrix, it is still
true that $P^{-1}gP$ diagonalizes $g$. Therefore, Proposition~2
guarantees that the equations of motion will decouple into a
2-dimensional system for $u$ and $v$ and a separate second-order
equation for $z$. Reversing the argument, this implies that
admissible scalar potentials for our original system necessarily
must be of the form
\begin{equation}
V(t,x) = U(t,u,v) + Z(t,z), \label{U+Z}
\end{equation}
$U$ and $Z$ being arbitrary functions of the indicated variables.
This concludes the analysis for the case that $\Gamma^1_2$ in the
relations (\ref{Gamma13-23}) was required to depend explicitly on at
least one of the $x$-variables.

What about the case that $\Gamma^1_2$ depends on time only, say
$\Gamma^1_2=\sigma(t)$? Using the representation (\ref{Gamma13-23}),
the matrix of connection coefficients then has the form
\begin{equation}
(\Gamma^a_b) = \left( \begin{array}{ccc} 0 & \sigma &
a\sigma+\tilde{a} \\ -\sigma & 0 & b\sigma+\tilde{b} \\ -
(a\sigma+\tilde{a}) & - (b\sigma+\tilde{b}) & 0 \end{array} \right).
\label{Gamma^a_b}
\end{equation}
As argued before, we can fix the admissible vector potentials,
without loss of generality, to be
\begin{equation}
A^a(t,x) = \Gamma^a_b(t) x^b, \label{An3}
\end{equation}
since the addition of gradient terms can be absorbed in the scalar
potential up to a total time-derivative. The curvature condition
(\ref{curv}) is automatically satisfied here but, unlike the
situation we encountered in our discussion of examples for $n=2$,
the remaining conditions (\ref{Veqn}) do not reduce to equations
involving the function $V$ only and in fact turn out to be quite
complicated. A different line of approach can be adopted for this
situation, which in fact applies to systems with arbitrary dimension
and is explained in the next section.

\section{The case of connection coefficients depending on time only}

The inspiration comes from the discussion on linear systems in
\cite{SEB82}: it is a well known practice for the integration of
systems of linear second-order equations to eliminate the terms
linear in the velocities by an appropriate coordinate
transformation. This idea applies just as well for non-linear
equations as long as they have linear dependence on the velocities.
If the $\Gamma^a_b$ are functions of time only, the equations we
start from, written with matrix notations, are
\begin{equation}
\ddot{x} + 2\Gamma(t)\dot{x} + \fpd{V}{x}(t,x) + \dot{\Gamma}(t)x=0.
\label{x-eqns}
\end{equation}
If $U(t)$ is the solution of the matrix differential equation
\begin{equation}
\dot{U} + \Gamma U =0, \qquad U(0)= I_n, \label{Ueqn}
\end{equation}
then $g(t)=(U^{-1})^T S U^{-1}$ is the solution of the equation
$\dot{g}=g\Gamma + (g\Gamma)^T$ with initial value $g(0)=S$, and the
coordinate transformation $x=U(t)y$ will eliminate the (linear)
velocity dependence in the forces. But remember that $\Gamma$ is
skew-symmetric in our case, so that (\ref{Ueqn}) implies that
$\dot{U}^T-U^T\Gamma=0$, from which it subsequently follows that
$U^T\dot{U}+ \dot{U}^TU=0$. Hence $U^TU$ is constant and since
$U(0)=I_n$, $U(t)$ will actually be an orthogonal matrix with the
identity matrix as initial value. It is further straightforward to
verify that the transformed differential equations for the
$y$-variables will read:
\[
\ddot{y} -(U^T\Gamma^2U)y + \fpd{\bar{V}}{y} =0,
\quad\mbox{with}\quad \bar{V}(t,y)= V(t,U(t)y),
\]
or
\begin{equation}
\ddot{y} + \fpd{W}{y}=0, \quad \mbox{with} \quad W(t,y)= \bar{V} -
\onehalf y^T(U^T\Gamma^2U)y. \label{y-eqns}
\end{equation}
Observe that the $y$-equations still have the unit matrix as
multiplier since the forces are of potential type. More
interestingly is the fact that the alternative multiplier $g(t)$ for
the original equations translates to the existence of a constant
alternative multiplier $S$ for the $y$-equations. What's more in
making this transition we have eliminated the problem that the
remaining conditions (\ref{Veqn}) had the complicating coupling
between scalar and vector potential, because they now simply read
\begin{equation}
S_{ac}\fpd{^2W}{y^c\partial y^b} = S_{bc}\fpd{^2W}{y^c\partial y^a}.
\label{Weqn}
\end{equation}

We summarize our results in the following statement.

\begin{prop} Assume that the \sode\ $\Gamma$ on $\R^{2n+1}$ admits a coordinate
representation for which $\delta_{ab}$ is a multiplier matrix for
the inverse problem of Lagrangian mechanics and that the connection
coefficients are functions of time only. Then, without loss of
generality, the vector potential can be taken to be of the form
$A^a(t,x) = \Gamma^a_b(t) x^b$, and for there to exist an
alternative multiplier, it suffices to construct a non-singular,
constant symmetric matrix $S$ (not a multiple of the identity) and a
function $W(t,y)$, satisfying the requirements (\ref{Weqn}). If
$U(t)$ is the solution of the matrix differential equation
(\ref{Ueqn}), the admissible scalar potentials for the given system
will be of the form
\begin{equation}
V(t,x) = W(t,U^T(t)x) + \onehalf x^T\Gamma^2(t)x, \label{admV}
\end{equation}
and the corresponding alternative multiplier is given by $g(t)=
U(t)S U^T(t)$. \qed
\end{prop}

It should be emphasized that the orthogonal transformation with the
matrix $U$ in this picture is different from the one with $P$ used
before: $P$ is dictated by the diagonalization of the multiplier
$g$, whereas $U$ is dictated by the matrix of connection
coefficients $\Gamma(t)$. In other words there is no reason why the
constant matrix $S$ in Proposition~3 (which is the initial value of
$g$) would be diagonal and therefore, the transformed equations
(\ref{y-eqns}) are different from the decoupled equations which
Proposition~2 guarantees to exist. One can easily combine the
results of both propositions: once an admissible system would have
been reduced to the form (\ref{y-eqns}) and we accordingly know the
initial value $S$ of the alternative multiplier $g$, it suffices to
diagonalize $S$ to find a further transformation which will decouple
the system. Conversely, starting from a number of decoupled systems,
each of the form $\ddot{z}_{(i)}= -\partial W_{(i)}/\partial z$ for
some scalar potential $W_{(i)}$, multiply each subsystem with an
appropriate constant $\lambda_i$ and then apply an arbitrary
constant, orthogonal transformation to the direct sum total system;
the result will be a potential system which has the property
(\ref{Weqn}), with $W=\sum \lambda_iW_{(i)}$, expressed in the new
$y$-variables. Next, select an arbitrary orthogonal matrix $U(t)$
which has the unit matrix as initial value and define $\Gamma(t)$ by
$\Gamma:=-\dot{U}U^T$. Then, appealing to the results of
Proposition~3, equations of motion of electromagnetic type with the
desired property should be of the form:
\begin{equation}
\ddot{x}^a = - 2\Gamma^a_c(t)\dot{x}^c - \fpd{V}{x^a} -
\fpd{A^a}{t}, \label{prop3-1}
\end{equation}
with
\begin{equation}
A^a=\Gamma^a_b(t)x^b, \qquad \mbox{and}\qquad V= W(t,U^T(t)x) +
\onehalf x^T\Gamma^2(t)x. \label{prop3-2}
\end{equation}

We illustrate this procedure by constructing a specific example for
$n=3$ by starting somehow in the middle of the converse procedure we
just explained, i.e.\ in the representation (\ref{y-eqns}). Consider
the function
\[
W(t,y) = a(t)(y_1- p\, y_2)^3, \qquad  \mbox{$p$ constant}.
\]
The general symmetric matrix $S$ for which (\ref{Weqn}) holds with
this $W$, is given by
\[
S = \left( \begin{array}{ccc} s_1 & k & mp \\
k & s_1 + (k/p)(1-p^2) & m \\ mp & m & s_3 \end{array} \right)
\]
but to avoid expression swell, let us make the following choice for
the five free constant parameters still at our disposal:
$s_1=s_3=k=p=1,\ m=0$. Obviously the $y$-equations we are taking as
starting point happen to exhibit a certain feature of decoupling
already, they read
\[
\ddot{y}_1 = - 3 a(t) (y_1- y_2)^2, \qquad \ddot{y}_2 = 3 a(t) (y_1-
y_2)^2, \qquad \ddot{y}_3 =0.
\]
We choose $U(t)$ to represent a simple rotation around de $y_1$
axis:
\[
U(t) = \left( \begin{array}{ccc} 1&0&0\\
0&\cos\theta(t)&-\sin\theta(t) \\ 0&\sin\theta(t)&\cos\theta(t)
\end{array}\right)
\]
then $\Gamma:=-\dot{U}U^T$ becomes
\[
\Gamma = \left( \begin{array}{ccc} 0&0&0\\ 0&0&\sigma \\
0&-\sigma&0\end{array} \right)
\]
where we have put $\sigma(t)=\dot{\theta}(t)$. All we need now to
write down admissible $x$-equations of the form (\ref{prop3-1}) is
the scalar potential $V(t,x)$: it is given by
\[
V(t,x) = a(t)\big(x_1 - (\cos\theta)x_2 - (\sin\theta) x_3\big)^3 -
\onehalf \sigma^2(x_2^2+x_3^2).
\]
It is clear even without writing them down that the $x$-equations
will be fully coupled. However, they have the alternative multiplier
\[
g= \left( \begin{array}{ccc} 1&\cos\theta&\sin\theta \\
\cos\theta&1&0 \\ \sin\theta&0&1 \end{array}\right).
\]
The matrix $g$ (just like $S$ of course) has three distinct
eigenvalues, namely $2, 0, 1$. Hence, Proposition~2 proclaims that
the $x$-equations should decouple completely in appropriate
coordinates (and of course the partially coupled $y$-equations we
started from just as well). It is a fairly straightforward exercise
(at least with some computer algebra assistance) to compute a matrix
$P(t)$ which diagonalizes $g(t)$ and to verify that the resulting
transformation $x=P(t)z$ does indeed have the predicted decoupling
effect. But note with regard to the $y$-equations that the
transition to the decoupling $z$-coordinates may have re-introduced
velocity terms, resulting from the fact that the matrix $P(t)$ which
diagonalizes $g$ need not be taken strictly orthogonal, and that
$P^{-1}U$ accordingly need not be constant. Of course, if we
diagonalize the constant multiplier $S$ of the $y$-equations via a
constant matrix, we obtain another transition to a decoupled
$z$-system without velocity terms.

\subsubsection*{Acknowledgements} This work is part of the IRSES
project GEOMECH (nr.\ 246981) within the 7th European Community
Framework Programme. TM is a Postdoctoral Fellow of the Research
Foundation – Flanders (FWO). OK acknowledges support by the grant
GACR 201/09/0981 of the Czech Science Foundation. Each author warmly
thanks the others for the hospitality of their institutions while
this paper was being written.


\begin{thebibliography}{99}

\bibitem{APST}
J.\,E.\ Aldridge, G.\,E.\ Prince, W.\ Sarlet and G.\ Thompson, An
EDS approach to the inverse problem in the calculus of variations,
{\it J.\ Math.\ Phys.\/} {\bf 47} (2006) 103508.

\bibitem{AT92}
I.\ Anderson and G.\ Thompson, The inverse problem of the calculus
of variations for ordinary differential equations, {\it  Mem.\
Amer.\ Math.\ Soc.\/} {\bf 473} (1992).

\bibitem{Buca}
I.\ Bucataru and M.\,F.\ Dahl, Semi-basic 1-forms and Helmholtz
conditions for the inverse problem of the calculus of variations,
{\it J.\ Geom.\ Mech.\/} {\bf 1} (2009) 159--180.

\bibitem{Buca2}
I.\ Bucataru and O.\ Constantinescu, Helmholtz conditions and
symmetries for the time dependent case of the inverse problem of the
calculus of variations, {\it J.\ Geom.\ Phys.\/} {\bf 60} (2010)
1710–-1725.

\bibitem{Ferrario}
C.\ Ferrario, G.\ Lo Vecchio, G.\ Marmo, G.\ Morandi and C.\ Rubano,
\emph{A separability theorem for dynamical systems admitting
alternative Lagrangian descriptions\/}, J.\ Phys.\ A: Math.\ Gen.,
\textbf{20} (1987) 3225--3236.

\bibitem{GM1}
J.\ Grifone and Z.\ Muzsnay, On the inverse problem of the
variational calculus: existence of Lagrangians associated with a
spray in the isotropic case, {\it Ann.\ Inst.\ Fourier\/} {\bf 49}
(1999) 1387--1421.

\bibitem{Olga}
O.\ Krupkov\'a, Variational metrics on $\R\times TM$ and the
geometry of nonconservative mechanics, {\it Math.\ Slovaca\/} {\bf
44} (1994) 315--335.


\bibitem{olgageoff}
O.\ Krupkov\'{a} and G.\,E.\ Prince, Second order ordinary
differential equations in jet bundles and the inverse problem of the
calculus of variations, in {\it Handbook of Global Analysis\/} D.\
Krupka and D.\ Saunders eds.\ (Elsevier 2008) 837--904.

\bibitem{MP}
E.\ Massa and E.\ Pagani, Jet bundle geometry, dynamical
connections, and the inverse problem of Lagrangian mechanics, {\it
Ann.\ Inst.\ H.\ Poincaré, Phys.\ Théor.\/} {\bf 61} (1994) 17–-62.

\bibitem{S82}
W.\ Sarlet, The Helmholtz conditions revisited. A new approach to
the inverse problem of Lagrangian dynamics, {\em  J.\ Phys.\ A:
Math.\ Gen.\/} {\bf 15} (1982) 1503--1517.

\bibitem{SEB82}
W.\ Sarlet, E.\ Engels and L.Y.\ Bahar, Time-dependent linear
systems derivable from a variational principle, \emph{Int.\ J.\
Engng.\ Sci.\/} {\bf 20} (1982) 55--66.

\bibitem{SP10}
W.\ Sarlet and G.\ Prince, Alternative kinetic energy metrics for
Lagrangian systems, {\em J.\ Phys.\ A: Math.\ Theor.\/} {\bf 43}
(2010) 445204.

\bibitem{SVCM95}
W.\ Sarlet, A.\ Vandecasteele, F.\ Cantrijn and E.\ Mart\'{\i}nez,
Derivations of forms along a map: the framework for time--dependent
second--order equations, {\em Diff. Geom. Applic.\/} {\bf 5} (1995)
171--203.

\end{thebibliography}
\end{document}